\documentclass[twocolumn]{aastex631}
\usepackage{amsmath}
\usepackage{enumitem}
\usepackage{booktabs}
\usepackage{makecell}
\newcommand{\mytilde}{\raise.19ex\hbox{$\scriptstyle\sim$}}
\newcommand{\qcr}[1]{{\fontfamily{qcr}\selectfont #1}}
\newcommand{\bs}{\boldsymbol}

\shorttitle{Weak-Lensing Analysis of the Complex Merging Cluster Abell 514}
\shortauthors{Ahn et al.}

\begin{document}


\title{Substructures within Substructures in the Complex Post-Merging System A514 Unveiled by High-Resolution Magellan/Megacam Weak Lensing}

\author[0009-0009-4676-7868]{Eunmo Ahn}
\affiliation{Department of Astronomy, Yonsei University, 50 Yonsei-ro, Seoul 03722, Korea; eunmo.ahn@yonsei.ac.kr, mkjee@yonsei.ac.kr}

\author[0000-0002-5751-3697]{M. James Jee}
\affiliation{Department of Astronomy, Yonsei University, 50 Yonsei-ro, Seoul 03722, Korea; eunmo.ahn@yonsei.ac.kr, mkjee@yonsei.ac.kr}
\affiliation{Department of Physics, University of California, Davis, One Shields Avenue, Davis, CA 95616, USA}

\author[0000-0002-1566-5094]{Wonki Lee}
\affiliation{Department of Astronomy, Yonsei University, 50 Yonsei-ro, Seoul 03722, Korea; eunmo.ahn@yonsei.ac.kr, mkjee@yonsei.ac.kr}

\author[0000-0001-9139-5455]{Hyungjin Joo}
\affiliation{Department of Astronomy, Yonsei University, 50 Yonsei-ro, Seoul 03722, Korea; eunmo.ahn@yonsei.ac.kr, mkjee@yonsei.ac.kr}

\author[0000-0003-3175-2347]{John ZuHone}
\affiliation{Harvard-Smithsonian Center for Astrophysics, 60 Garden St., Cambridge, MA 02138, USA}

\begin{abstract}
Abell 514 (A514) at $z=0.071$ is an intriguing merging system exhibiting highly elongated ($\mytilde1$~Mpc) X-ray features and three large-scale ($300\sim500$ kpc) bent radio jets.
To dissect this system with its multi-wavelength data, it is critical to robustly identify and quantify its dark matter (DM) substructures.
We present a weak-lensing analysis of A514 using deep Magellan/Megacam observations.
Combining two optical band filter imaging data obtained under optimal seeing ($\mytilde0\farcs6$) and leveraging the proximity of A514, we achieve a high source density of $\mytilde46~\mbox{arcmin}^{-2}$ or $\mathrm{\mytilde 6940~ Mpc^{-2}}$, which enables high-resolution mass reconstruction.
We unveil the complex DM substructures of A514, which are characterized by the NW and SE subclusters separated by $\mytilde0.7$~Mpc, each exhibiting a bimodal mass distribution.
The total mass of the NW subcluster is estimated to be $\mathrm{M^{NW}_{200c} = 1.08_{-0.22}^{+0.24} \times 10^{14} M_{\odot}}$ and is further resolved into the eastern  ($\mathrm{M^{NW_E}_{200c} = 2.6_{-1.1}^{+1.4} \times 10^{13} M_{\odot}})$ and western ($\mathrm{M^{NW_W}_{200c} = 7.1_{-2.0}^{+2.3} \times 10^{13} M_{\odot}}$) components.
The mass of the SE subcluster is $\mathrm{M^{SE}_{200c} = 1.55_{-0.26}^{+0.28} \times 10^{14} M_{\odot}}$, which is also further resolved into the northern ($\mathrm{M^{SE_N}_{200c} = 2.9_{-1.3}^{+1.8} \times 10^{13} M_{\odot}}$) and southern ($\mathrm{M^{SE_S}_{200c} = 8.5_{-2.6}^{+3.0} \times 10^{13} M_{\odot}}$) components.
These four substructures coincide with the A514 brightest galaxies and are detected with significances ranging from 3.4$\sigma$ to 4.8$\sigma$.
Comparison of the dark matter substructures with the X-ray distribution suggests that A514 might have experienced an off-axis collision, and the NW and SE subclusters are currently near their apocenters. 
\end{abstract}


\section{Introduction}\label{sec:introduction}
According to the hierarchical structure formation paradigm,
mergers are one of the most important channels for the growth of galaxy clusters.
Although the mergers are primarily driven by the dark matter halos, the dominant source of the gravitational potential, the results leave several distinct features on their hot X-ray emitting intracluster medium (ICM), comprising
$10-15$\% of the total cluster mass budget.
These features are identified as ICM-galaxy dissociation, cold fronts, shocks, radio relics, radio halos, bent radio jets, etc., in multi-wavelength observations \citep[e.g.,][]{markevitch2002,vanWeeren2010,cassano2010, Wonki.2023b}. Unfortunately, we still do not understand the detailed astrophysical processes giving rise to these characteristic multi-wavelength features.

To advance, numerical simulations are employed. This entails meticulous reconstruction of the merging scenarios and establishment of initial conditions.
The tasks are non-trivial because we can only access single snapshots in multi-Gyr-long cluster mergers. Among these, one of the most critical is the identification of dark matter substructures and their associated masses, as the merger trajectories are primarily driven by gravity.
The observed distribution of ICM by X-ray cannot  reliably indicate the dark matter substructures because the plasma particles interact through Coloumb forces, and their distributions can differ significantly from those of corresponding dark matter subhalos. The galaxy distributions can be considered as indicators of dark matter substructures. They may serve as better indicators than the ICM because galaxies are effectively collisionless during the merger. However, galaxies are only biased tracers of mass, sparsely sampling the dark matter subhalos. Most importantly, it is paramount to properly quantify the subhalo masses, as well as the subhalo positions, since as mentioned above, they predominantly influence the merger trajectories. Neither the plasma nor the galaxy distribution is a dependable proxy of the subhalo masses.

Weak gravitational lensing (WL hereafter) is a powerful tool for mapping the dark matter distribution of galaxy clusters.
Since it probes the projected mass solely based on the gravitational lensing effects on the shapes of background galaxies, WL does not necessitate assumptions about the dynamical state of the lens.
In particular, merging clusters are believed to deviate substantially from the hydrostatic equilibrium, making this merit even more critical.
Rich dark matter substructures have been detected by WL in various merging clusters \citep[e.g.,][]{Okabe.2008, Jee2009, Jee.2012, Wittman.2014, Martinet.2016, Finner.2017}.

In this study, we present the first WL analysis of the low-redshift ($z=0.071$) merging galaxy cluster Abell 514 (hereafter A514).
One prominent feature of A514 is its extended X-ray emission stretching $\mytilde1$~Mpc from northwest (NW) to southeast (SE), which connects two galaxy overdensities.
With XMM-Newton observations, \cite{Weratschnig.2008} claimed the detection of an ICM density discontinuity and suggested an ongoing NW-SE merger.
Another intriguing characteristic of A514 is its peculiar radio features, including three head-tail radio galaxies with bent morphologies \citep[e.g.,][]{Burns.1994, Govoni.2001}.
Remarkably, the radio emission originating from the two radio lobes of the AGN in the SE region extends $\mytilde0.7$~Mpc towards the southern outskirts with multiple bends \citep{Wonki.2023b}.
Although there is no direct evidence to date, it is highly probable that these peculiar radio features are influenced by the ongoing merger.
One of the goals of the current WL study is to identify and quantify the substructures in A514, which will provide critical input to future numerical simulations.
An outstanding question in merging cluster physics is how AGN plasma gets redistributed within the ICM due to merger-driven gas motions \citep[see][for review]{vazza2024}.

WL analysis of low-redshift clusters has been considered a challenge relative to intermediate redshift ($z\sim0.5$) clusters because of their low lensing efficiency (i.e., smaller distortion given the same mass and source redshift).
However, this disadvantage is outweighed by the large projected area thanks to their proximity, which provides a significantly higher number of background galaxies per physical area at the lens redshift.
For instance, \cite{HyeongHan.2024} demonstrated that while at $z=0.02$ the lensing efficiency is lower by an order of magnitude than at $z=0.5$, the net gain in S/N per physical area is approximately three times higher.
Thus, the low-redshift merging cluster WL provides an excellent  opportunity to probe the complex substructures in great detail.
We note, however, that since the intrinsic lensing efficiency is low, the requirement for systematics control is high for low-redshift WL.

Throughout this paper, we assume a flat $\mathrm{\Lambda CDM}$ cosmology with $H_0\mathrm{=70~km~s^{-1}Mpc^{-1}}$ and $\mathrm{\Omega_m=0.3}$.
At the cluster redshift ($z=0.071$), the angular size of $1'$ corresponds to the physical size of $\mathrm{\mytilde 82~ kpc}$.
$\mathrm{M_{200c}}$ ($\mathrm{M_{500c}}$) is defined as the mass enclosed by a sphere inside which the average density equals to 200 (500) times the critical density at the cluster redshift. All errors are quoted at the 1-$\sigma$ level unless otherwise noted.

\section{Observation and Data Reduction}\label{sec:observation and data reduction}
The A514 field was observed using the Magellan/Megacam imager on the night of October 23, 2022, in the $g$- and $r$- bands (PI: W. Lee).
The Megacam imager's focal plane is composed of 36 CCDs, yielding a total field of view of $25' \times 25'$ \citep{McLeod.2015}.
We applied dithering and field rotation among all 18 pointings for both $g$- and $r$- filters.
This significantly reduces several artifacts that negatively affect WL analysis around bright stars, including their diffraction spikes and saturation trails.
Consequently, the number of usable source galaxies for WL increases.
The total exposure for each filter is 5400 s.
The mean seeings of the $g$- and $r$- filters are  $0.\!''61$ and $0.\!''60$, respectively, which are ideal for ground-based WL.

We applied initial bias and sky-flat correction, and cosmic rays were masked using {\tt ASTROSCRAPPY} \citep{McCully.2018}.
All flat-fielded frames underwent processing by {\tt SExtractor} \citep{Bertin.1996} for the preparation of the astrometric and photometric calibration using {\tt SCAMP} \citep{Bertin.2006}.
The final deep mosaic images where the weak-lensing signal is measured were created with {\tt SWARP} \citep{Betrin.2002} by stacking all the frames precisely with the refined World Coordinate System (WCS) information.

In general, mean-stacked images provide a better S/N than the median-stacked images.
However, a plain mean-stacking scheme is vulnerable to outliers.
For this reason, we used the {\tt SWARP} keyword \qcr{COMBINE\_TYPE=CLIPPED} \citep{Gruen.2014} for outlier-clipped mean stacking.
This option first generates a median-stacked image and then produces the final image through inverse-variance weight-averaging, clipping the outliers that deviate significantly from the median-stacked image.
Additionally, this option generates a log file containing information about photometric outliers in each frame.
This information is essential for later PSF modeling, where the appropriate weight needs to be applied.
Readers are referred to \cite{Gruen.2014} and \cite{Jee.2015} for more details.
We found that the background rms is about $\mathrm{\mytilde11\%}$ lower in the mean-stacked image than in the median-stacked image, both for $g$- and $r$- bands.
Although the overall flux level is higher in the $r$-band, the photometric S/N for faint galaxies is slightly higher in the $g$-band.
Therefore, we conducted WL analysis using both $g$- and $r$-bands.

We executed {\tt SExtractor} in dual-image mode, where the detection image was generated by weight-averaging the $g$- and $r$- band mosaic images.
The dual-image mode maintains consistent isophotal area across all filters, ensuring that object colors were measured from identical isophotal apertures.
We defined objects as regions with more than five connected pixels, each having a flux above 1.5 times the local background rms.


\section{Weak Lensing Analysis}\label{sec:weak lensing analysis}
\begin{figure*}
    \centering
    \includegraphics[width=\columnwidth]{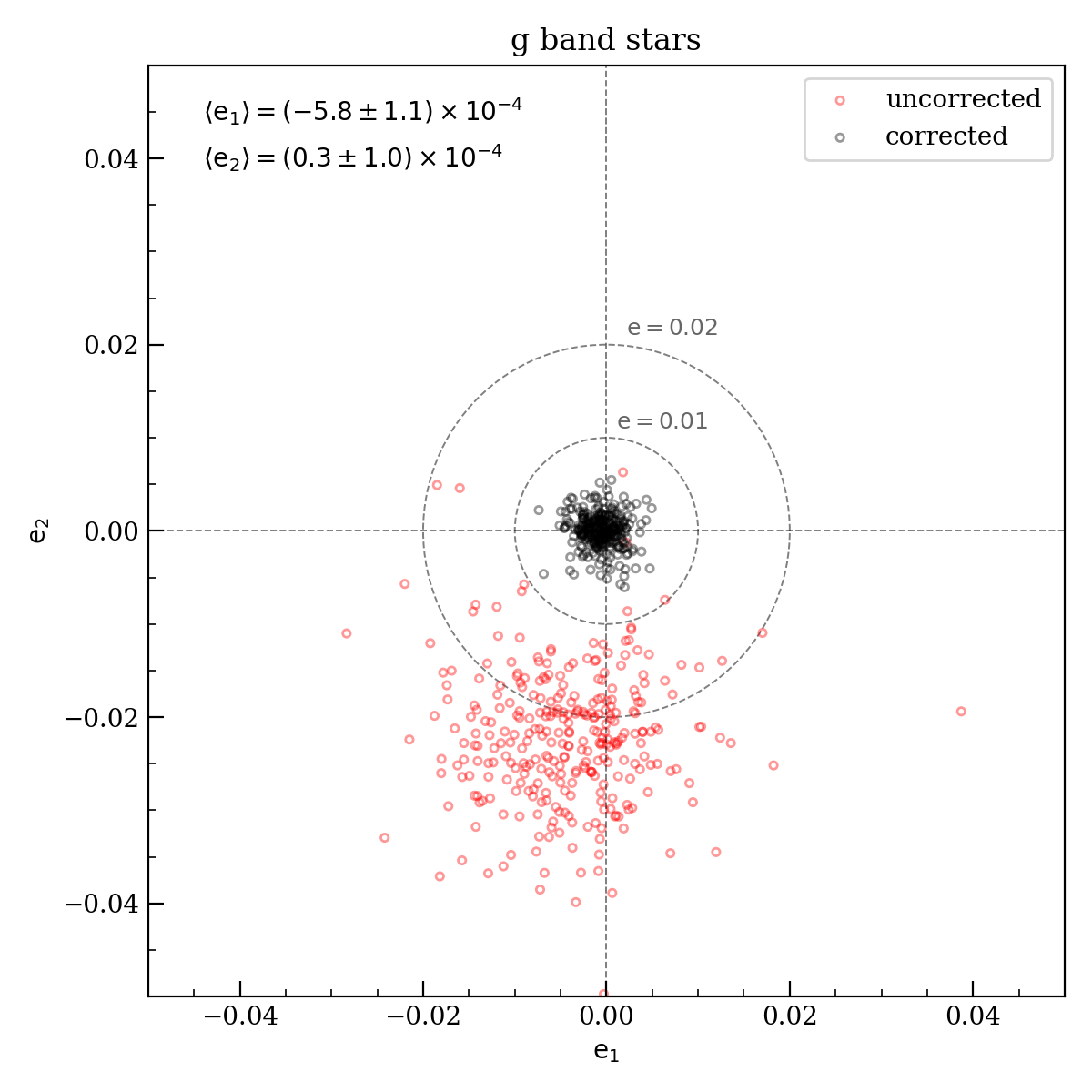}
    \includegraphics[width=\columnwidth]{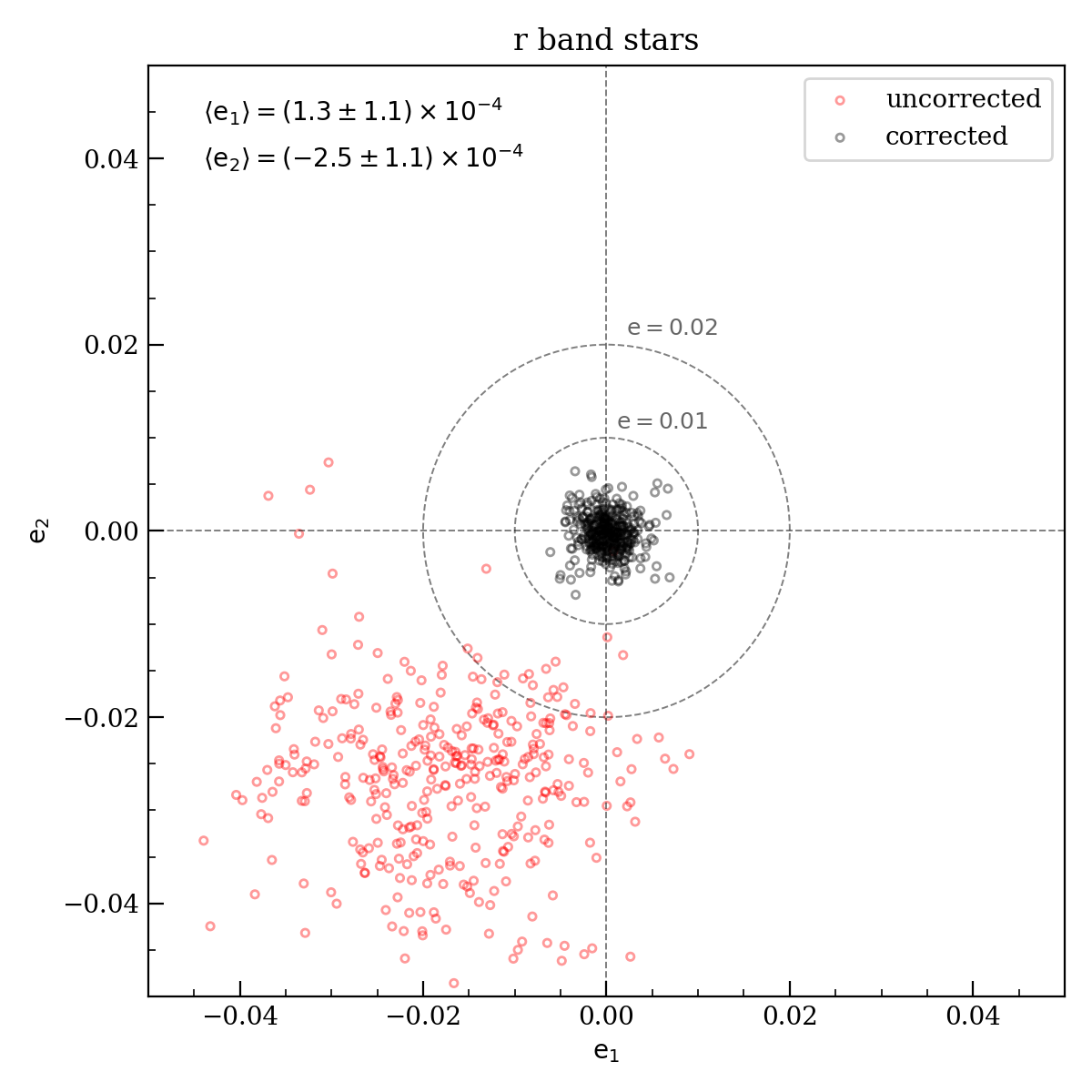}
    \caption{Megacam PSF modeling for $g$-band (left) and $r$-band (right) images. Red circles represent the ellipticity components of the observed stars in the mosaic image. Black circles represent the ellipticity components of the residuals between the observed stars and the modeled PSF. After correction, the residuals in both $g$- and $r$-bands are centered at the origin with reduced scatter, ensuring the precise and accurate PSF modeling.}
    \label{fig:psf ellipticity conponent}
\end{figure*}

\subsection{Basic Theory}\label{sub:basic theory}
Measuring the local ensemble-averaged distortion of background galaxies caused by the lens allows us to investigate the dark matter distribution and mass of the galaxy cluster.
In this section, we provide a brief overview of the basic theory and formalism of WL.
Readers are referred to review papers \citep[e.g.,][]{Narayan.1996, Bartelmann.2001, Schneider.2006} for details.

Light bundles from background galaxies bend under the gravitational tidal field caused by a foreground cluster.
The coordinate mapping from the background source plane $\bs{\beta}$ to the observed image plane $\bs{\theta}$ can be described by the lens equation:
\begin{equation}\label{eq:lens equation}
    \bs{\beta} = \bs{\theta} - \bs{\alpha}(\bs{\theta}),
\end{equation}
where $\bs{\alpha}$ is the scaled deflection angle, which involves the convergence $\kappa$ via:

\begin{equation}
    \bs{\alpha}(\bs{\theta}) = \frac{1}{\pi}\int_{\mathbb{R}^2} d^2 \bs{\theta '} \kappa(\bs{\theta'}) \frac{\bs{\theta} - \bs{\theta'}}{|\bs{\theta} - \bs{\theta'}|^2}.
\end{equation}
The convergence $\kappa$ is a dimensionless quantity defined by the projected surface mass density ($\Sigma$) 
normalized by the critical surface density ($\Sigma_c$):
\begin{equation}\label{eq:convergence}
    \kappa = \frac{\Sigma}{\Sigma_c}, \;\; \Sigma_c = \frac{c^2}{4 \pi G} \frac{D_s}{D_l D_{ls}},
\end{equation}

where $c$ is the speed of light, $G$ is the gravitational constant, and $D_s$, $D_l$, and $D_{ls}$ denote the angular diameter distances from the observer to the source, from the observer to the lens, and from the lens to the source, respectively.

If the size of a source is much smaller than the angular scale over which the lens properties vary, the mapping relation can be linearized with the Jacobian matrix A:
\begin{equation}\label{eq:jacobian}
    \bs{A} = \frac{\partial\bs{\beta}}{\partial\bs{\theta}} = (1-\kappa)\begin{pmatrix} 1-g_1 & -g_2 \\ -g_2 & 1+g_1 \end{pmatrix},
\end{equation}

where $g$ is the reduced shear, defined as

\begin{equation}\label{eq:reduced shear}
    g = \frac{\gamma}{1 - \kappa}.
\end{equation}
In Equation (\ref{eq:reduced shear}), $\gamma$ is the shear, which is directly related to the mass of the lens.
In the weak-lensing regime ($\kappa \ll 1$), the reduced shear $g$ approaches the shear $\gamma$.

We can use the complex notation for the reduced shear, introducing its two components:

\begin{equation}
    \bs{g} = g_1 + \bs{i}g_2.
\end{equation}
Here, $g_1$ distorts the image along the $x$-axis and $y$-axis directions, while $g_2$ distorts the image along the $y=x$ and $y=-x$ directions. 
In the same way, we can define the intrinsic (unlensed) ellipticity $\bs{\epsilon}$ and observed (lensed) ellipticity $\bs{e}$ using the same complex notation:

\begin{equation}
    \bs{\epsilon} = \epsilon_1 + \bs{i} \epsilon_2, \; \; \;
    \bs{e} = e_1 + \bs{i} e_2.
\end{equation}
Then, with the presence of the local reduced shear $\bs{g}$, the Equation (\ref{eq:jacobian}) will transform the intrinsic ellipticity $\bs{\epsilon}$ to the observed ellipticity $\bs{e}$ as:

\begin{equation}\label{eq:ellipticity transform}
    \bs{e} = \frac{\bs{\epsilon} + \bs{g}}{1 + \bs{g}^* \bs{\epsilon}},
\end{equation}
where the asterisk (*) denotes the complex conjugate.
Assuming that the intrinsic orientation of background galaxies is random, i.e. $\left\langle \bs{\epsilon} \right\rangle = 0$, Equation (\ref{eq:ellipticity transform}) becomes:

\begin{equation}\label{eq:e average g}
    \bs{g} = \left\langle \bs{e} \right\rangle.
\end{equation}
Therefore, we can use $\left\langle \bs{e} \right\rangle$ as an unbiased estimator of the reduced shear for the weak distortions in an ideal case where no systematic bias is present.

\subsection{PSF Modeling}\label{sub:psf modeling}
Observed galaxy shapes are affected by point spread functions (PSFs), and this effect becomes larger for the ground-based observations because of atmospheric turbulence.
This phenomenon introduces a systematic bias in the shear measurement, making Equation (\ref{eq:e average g}) invalid.
Therefore, precise and accurate PSF modeling is a critical step in WL signal detection.
We adopted the principal component analysis (PCA) technique \citep{Jee.2007, Jee.2011} for our PSF modeling.

In the mean-stacked image, the PSF is equivalent to a linear combination of the PSFs from all contributing individual frames, with the same weight used in the mosaic image stacking stage.
Since the PSF pattern has discontinuities across the CCD boundaries \citep[e.g.,][]{Jee.2013}, our previous WL studies modeled PSFs CCD by CCD and exposure by exposure \citep[e.g.,][]{Finner.2021, Cho.2022}.
However, the A514 field does not provide a sufficient number of stars suitable for modeling the PSF within each CCD.
Fortunately, we found that for a given exposure, the star ellipticity pattern flows smoothly across the CCD gaps when we put all 36 CCDs together, although the global pattern changes exposure by exposure.
Therefore, in this study, we chose to model the PSF for the entire focal plane (all 36 CCDs) exposure by exposure and stack the models to create the final PSF model of the mosaic image.

We initially identified ``good stars'' in each exposure, which satisfy several criteria.
First of all, these stars should be isolated without being saturated.
And they should not be located in the vicinity of the CCD boundaries.
Also, their centroid measured by the peak positions (\qcr{XPEAK\_IMAGE, YPEAK\_IMAGE}) should agree with the windowed centroid (\qcr{XWIN\_IMAGE, YWIN\_IMAGE}) within 1 pixel.
We were able to select about $\mytilde200$ ``good stars'' from each exposure for both $g$- and $r$-band images.
Then, all surviving stars were cropped into 21 pixel $\times$ 21 pixel postage-stamp images and used for PCA.
To create a PSF model in the mosaic image, we performed an inverse-variance weight-averaging of all contributing PSF models from individual exposures.
As mentioned earlier, these weights are identical to those applied in the creation of the mosaic image.

Figure \ref{fig:psf ellipticity conponent} shows the results from the PSF modeling in $g$- and $r$- bands.
The centroids of the residual ellipticities (the difference between the model and observed PSF pattern) are well-clustered around the origin.
Both the reduction in scatter and the shift toward the origin (0,0) confirm that our PSF model on the mosaic image is precise and accurate.

\subsection{Multiple-filter Shape Measurement}\label{sub:shape measurement}
We used a PSF-convolved elliptical Gaussian function to fit the light profile of background galaxies.
Although a $\mathrm{S\acute{e}rsic}$ model provides a better description, we find that the resulting shape measurement is noisier since the fitting assigns higher weights to the noisy pixels in the peripheral region \citep{Jee.2013}, reducing the number of usable galaxies.

The PSF-convolved elliptical Gaussian function $M(x, y)$ is:
\begin{equation}
    M(x, y) = G(x, y) \otimes P(x, y),
\end{equation}
where $P(x, y)$ is the model PSF at the source position and $G(x, y)$ is the elliptical Gaussian function:
\begin{equation}
\begin{aligned}
    G(x, y) = I_{bg} + I_{pk} \exp \left[
    -\frac{(\Delta x \cos \theta + \Delta y \sin \theta)^2}{2\sigma_x^2} \right. \\
    \left. -\frac{(-\Delta x \sin \theta + \Delta y \cos \theta)^2}{2\sigma_y^2} \right],
\end{aligned}
\end{equation}
where $I_{bg}$ and $I_{pk}$ denote the background level and maximum flux level, respectively.
The $\Delta x$ and $\Delta y$ represent the distances from the centroid to each pixel $(x, y)$, respectively, $\sigma_x^2$ and $\sigma_y^2$ are the variances, and $\theta$ is the position angle measured counterclockwise from the $x-$axis.
We fixed the background level and centroid using the {\tt SExtractor} measurements \qcr{BACKGROUND} and (\qcr{XWIN\_IMAGE, YWIN\_IMAGE}), respectively.
By reducing the free parameters in $G(x, y)$ from seven to four, we can further reduce the measurement error, thereby increasing the source density of usable galaxies in our analysis.

We used the {\tt MPFIT} code \citep{Markwardt.2009} to obtain the best-fit model that minimizes the difference between the model $M(x, y)$ and the observed galaxy profile $O(x, y)$.
{\tt MPFIT} performs non-linear least squares minimization based on the Levenberg–Marquardt algorithm.
With the best-fit parameters, it also returns their 1-sigma errors computed from the Hessian matrix.
The free parameters are $I_{pk}$, $e_1$, $e_2$, and $b$, where $e_1$ and $e_2$ are the two ellipticity components, and $b$ is the semi-minor axis.
We defined the $e_1$ and $e_2$ as:
\begin{equation}
\begin{split}
    e_1 = e \cos(2\theta), \\
    e_2 = e \sin(2\theta),
\end{split}
\end{equation}
where ellipticity $e$ is defined using semi-major axis $a$ and semi-minor axis $b$ as $e=(a-b)/(a+b)$.

We cropped each source galaxy into a square postage stamp image, which constructs the $\chi^2$ function.
We combined the two $\chi^2$ functions for $g$- and $r-$band images and obtained one best-fit model that minimizes the total $\chi^2$ function as follows:
\begin{equation}\label{eq:chi2}
\begin{aligned}
    f = \chi_g^2 + \chi_r^2 = \sum_{i} \frac{[O_g(x, y)_i - M_g(x, y)_i]^2}{\sigma_{g, i}^2} \\ + \frac{[O_r(x, y)_i - M_r(x, y)_i]^2}{\sigma_{r, i}^2},
\end{aligned}
\end{equation}
where $O (x, y)_i$ and $M(x, y)_i$ indicate the observed and modeled galaxy postage stamp images, respectively, and $\sigma_i$ is the background rms noise.
The summation over $i$ represents the summation of all pixels belonging to the square postage stamp image for each galaxy.
The two models $M_g(x, y)$ and $M_r(x, y)$ share the same shape parameters ($e_1$, $e_2$, $b$) except for the peak intensities. 

\begin{figure}
    \includegraphics[width=\columnwidth]{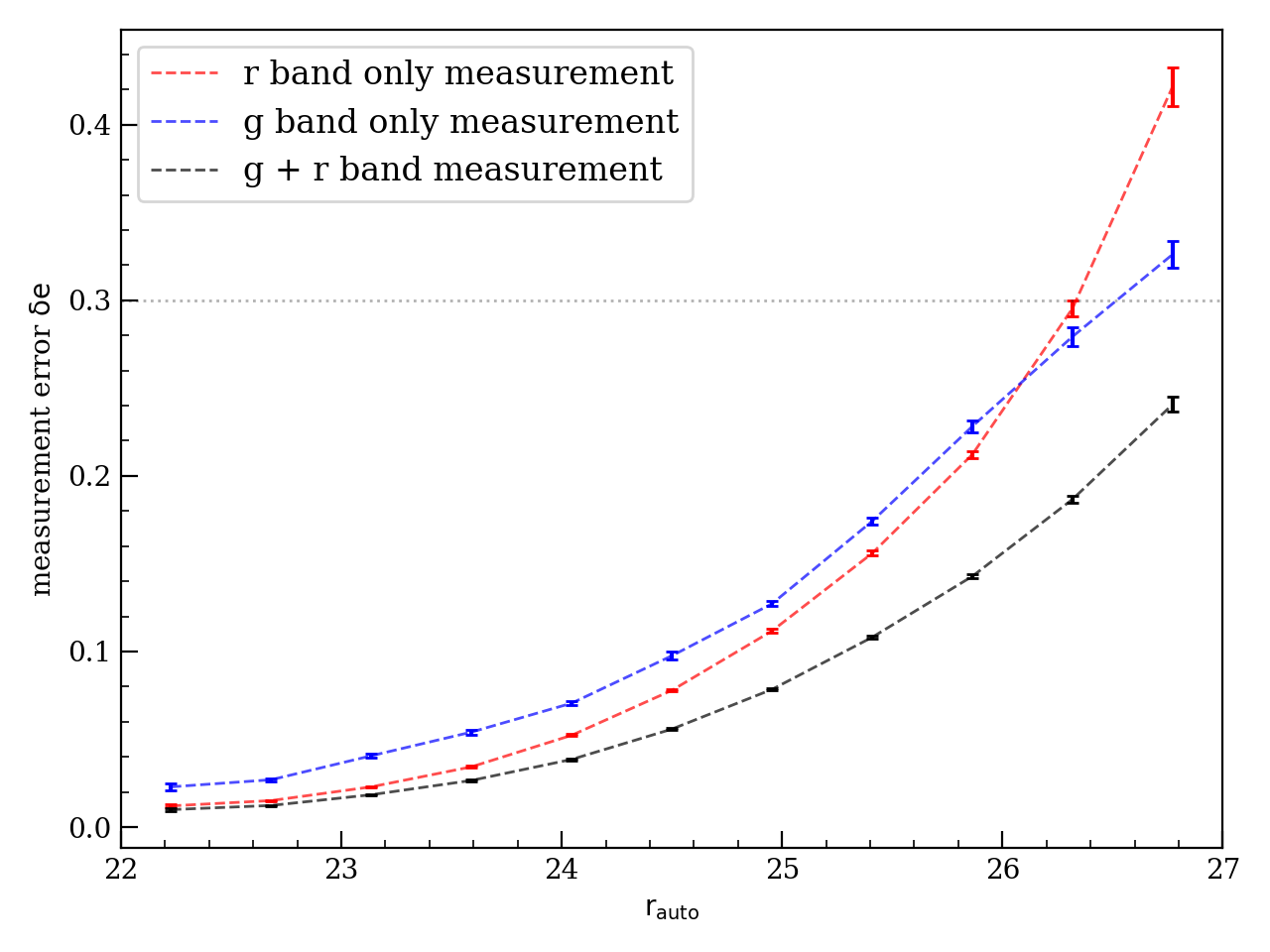}
    \caption{Ellipticity measurement errors determined from {\tt MPFIT} as a function of $r-$band magnitude. The measurement error for each case is the average of the $e_1$ and $e_2$ measurement errors. Across all magnitude ranges, especially for faint sources, combining two filters in the shape measurement decreases the measurement error. We obtained $\mytilde20\%$ more sources through simultaneous fitting given the same error cut (dotted line, $\delta e = 0.3$).}
    \label{fig:e measurement error}
\end{figure}

Figure \ref{fig:e measurement error} shows the measurement errors of ellipticity components as a function of $r$-band magnitude.
We compared three cases: two involving the fitting processes using only the $g$- or $r$-band image and the other using both $g$- and $r$- bands simultaneously.
Measurement errors from simultaneous fitting were consistently lower than those from single-band-only fitting across all magnitude ranges, and the difference is larger for fainter sources.
Consequently, extracting shape information from multiple filters reduces the measurement error, thereby increasing the total number of usable galaxies given the same measurement error cut.
In this study, we were able to obtain $\mytilde20\%$ more sources through simultaneous fitting.

A potential concern is the difference in shape between filters.
Detailed morphologies of galaxies vary from ultraviolet to infrared \citep[e.g.,][]{Kuchinski.2000}.
However, we used two broadband optical filters with a relatively small separation in effective wavelength.
As a result, the systematic difference in intrinsic ellipticity between the two filters is expected to be smaller than the measurement error \citep[e.g.,][]{Lee.2018}.

The analytic galaxy profiles cannot be the perfect representation of the real galaxies.
This model bias can lead to systematic errors in the shear measurement. Also, non-linear relations between pixel and ellipticity contribute to the systematic errors (noise bias).
By running image simulations that match observational data, we derived a global multiplicative factor of $m=1.14$ with a negligible ($ < \mytilde 10^{-3}$) additive bias \citep{Jee.2016}.
Readers are referred to \cite{Jee.2011} and \cite{Jee.2013} for details.

\subsection{Source Selection}\label{sub:source selection}
Optimal selection of background sources could be made by choosing objects whose photometric redshifts are greater than that of the cluster.
However, this is not the case for our data, and we selected WL sources based on their color-magnitude-redshift relation.

\begin{figure}
    \includegraphics[width=\columnwidth]{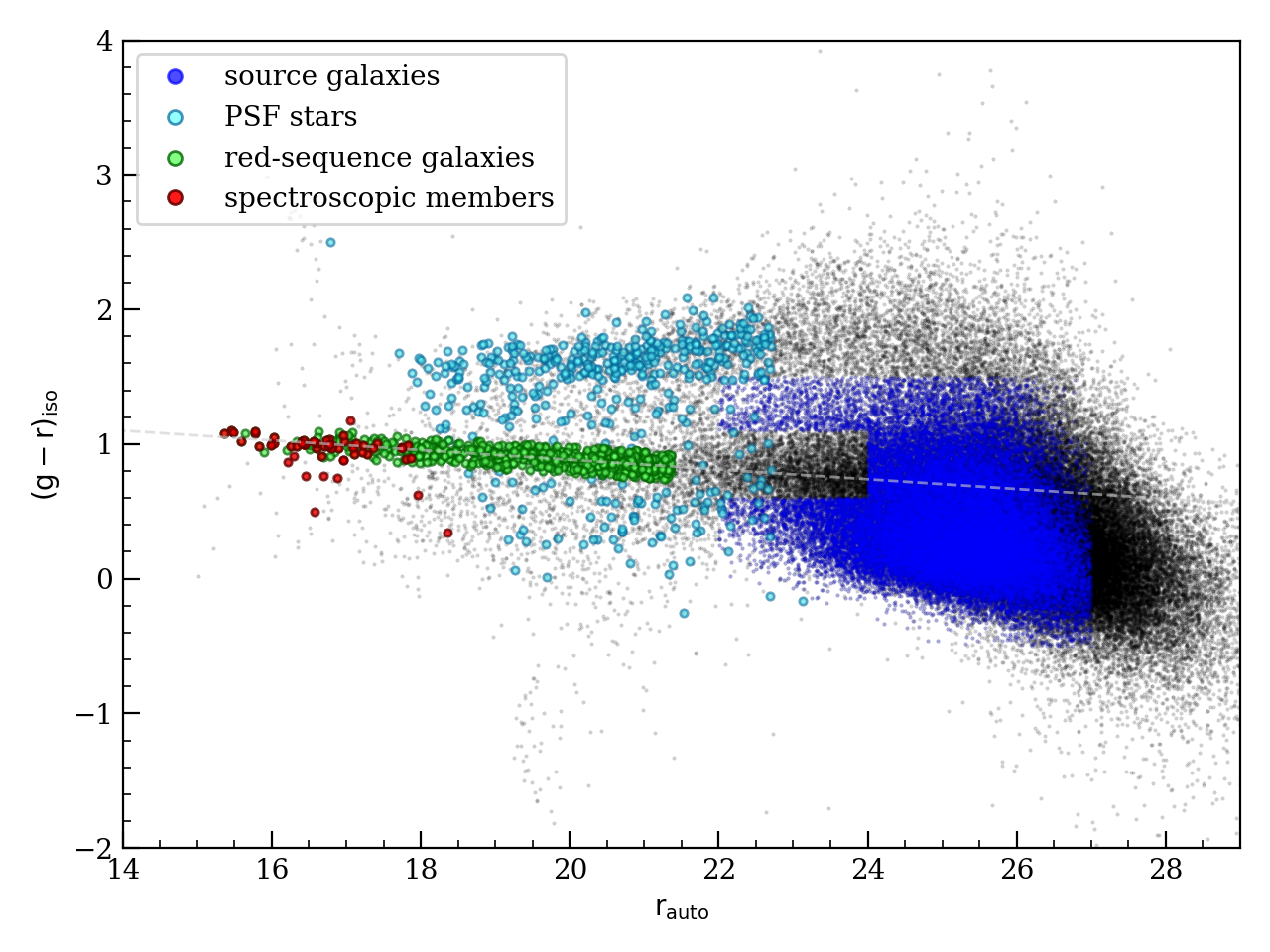}
    \caption{Color-magnitude diagram of the A514 field. Red dots represent spectroscopically confirmed members from the NASA/IPAC Extragalactic Database (NED) survey, and blue dots represent the background sources after applying source criteria. Green dots represent photometric member candidates selected through the red sequence (dashed-gray line). Many of the PSF stars, denoted as cyan dots, exhibit the horizontal locus redder than the cluster red sequence.}
    \label{fig:CMD}
\end{figure}

Figure {\ref{fig:CMD}} illustrates the color-magnitude diagram (CMD) of the A514 field. 
In this diagram, the red-sequence galaxies of A514 exhibit clear and tight color-magnitude relation.
Typically, for weak-lensing analysis of intermediate redshift ($z\sim0.5$) clusters, background sources are often chosen from a population fainter and bluer than the cluster red sequence.
However, given that A514 is located at $z \sim 0.07$, a majority of galaxies sufficiently fainter than the cluster red sequence are likely behind the cluster regardless of their colors.
We selected sources with $r-$band magnitudes in the range $22 < r_{auto} < 27$ and isophotal ($g - r$) colors within $-0.5 < (g-r)_{iso} < 1.5$.
For the galaxies whose colors match those of the A514 red sequence, we increase the magnitude upper limit to $r_{auto} = 24$ to minimize possible contamination from the faint end of the A514 red sequence.
The upper bound of the color is to avoid contamination from the Milky Way stars, which show distinct horizontal locus above the A514 red sequence as shown in Figure \ref{fig:CMD}.
We estimated that only $\mytilde4\%$ of the resulting source population is in the foreground by applying the same source criteria to the COSMOS photo-$z$ catalog.

As a measure to exclude spurious sources whose shape measurements are unreliable, the following additional criteria were applied:
\begin{enumerate}
\item The $\chi^2$ minimization should be reliable ({\tt MPFIT STATUS} = 1),
\item The semi-minor axis $b$ should be larger than 0.3 pixels, as shapes with smaller values show the ``bug pattern" reported in \cite{Jee.2013} because of pixelation issues regardless of their S/N,
\item The semi-major axis $a$ should be smaller than 30 pixels,
\item The total ellipticity ($e = \sqrt{e_1^2 + e_2^2}$) should be less than 0.9,
\item Measurement errors for both $e_1$ and $e_2$ should be less than 0.3, and
\item The {\tt SExtractor flag} should be less than 4 to exclude potentially problematic detections.
\end{enumerate}
Finally, through visual inspection, non-astrophysical sources, such as diffraction spikes or rings near bright stars, were discarded.
The final source catalog contains a total 45,353 galaxies, which provides a mean source density of $\mathrm{\mytilde 46.3~arcmin^{-2}}$ ($\mathrm{\mytilde 6940~Mpc^{-2}}$ at the A514 redshift).
Since the contribution from each source depends on its measurement error, we can define an effective WL source density \citep{Jee.2014a} as follows:
\begin{equation}\label{eq:eff num}
    \mathrm{n_{eff}} = \mathrm{\sum{\frac{\sigma_{SN}^2}{\sigma_{SN}^2 + (\delta e_i)^2}}},
\end{equation}
where $\mathrm{\sigma_{SN}}$ is the dispersion of the source ellipticity distribution per component ($\mytilde0.25$), and $\delta e_i$ is the $i$th galaxy's ellipticity measurement error per component.
From this equation, we estimate the effective source density to be $\mathrm{\mytilde 40.6~arcmin^{-2}}$.

\subsection{Redshift estimation}\label{sub:redshift estimation}
A weak-lensing signal is proportional to the angular diameter distance ratio $\beta = D_{ls} / D_s$ (Eqn~\ref{eq:convergence}).
We estimated the source redshift by employing the COSMOS2020 photometric redshift catalog \citep{Weaver.2022}.
As the catalog does not provide magnitudes in the Megacam photometric system, we used the Subaru/Hyper Suprime-Cam $g$- and $r$-magnitudes as proxies for our Megacam $g$- and $r$-magnitudes, respectively.
We applied the same color-magnitude criteria to the COSMOS2020 catalog, weighting the COSMOS2020 galaxies with the number density ratio in each magnitude bin between the COSMOS2020 and A514 fields \citep[e.g.,][]{Finner.2017, Mincheol.2019}.
This weighting scheme accounts for the difference in depth between the two fields.
Assigning zero weights to galaxies with redshifts smaller than that of A514, we calculated the effective $\beta$:
\begin{equation}
    \left\langle \beta \right\rangle = \left\langle max\left(0, \frac{D_{ls}}{D_s}\right) \right\rangle.
\end{equation}
We obtained $\left\langle \beta \right\rangle = 0.865$, which corresponds to an effective source redshift of $\mathrm{z_{eff}=0.607}$.
The assumption that all sources are located at a single redshift introduces bias because of the non-linearity in $\beta$.
To address the issue, we applied a first-order correction  \citep{Seitz.1997, Hoekstra.2000} as follows:
\begin{equation}\label{eq:first-order-correction}
    \frac{g'}{g} = 1 + \left(\frac{\left\langle \beta^2 \right\rangle}{\left\langle \beta \right\rangle^2}-1 \right) \kappa,
\end{equation}
where $g'$ and $g$ are the observed and true reduced shear, respectively.
We obtained $\left\langle \beta^2 \right\rangle = 0.781$, which scales the observed shear $g'$ by a factor of $(1 + 0.044\kappa)$.

\section{Results}\label{sec:results}

\begin{figure*}
    \centering
    \includegraphics[width=0.85\textwidth]{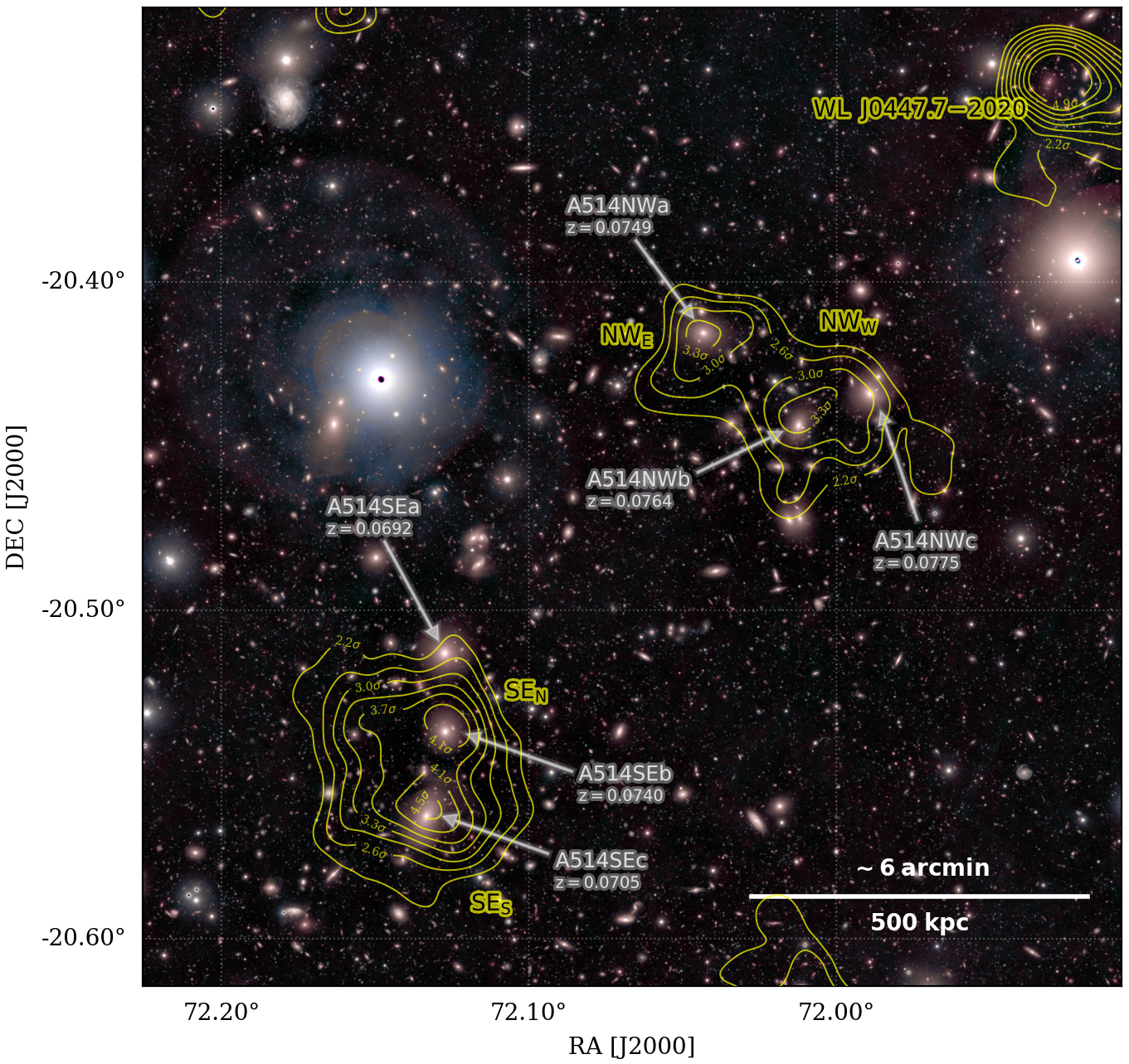}
    \caption{Convergence contours of the A514 field over a Magellan/Megacam color composite image. The contour levels represent statistical significance (see \textsection\ref{sub:centroid significances}). 
    On a large scale, A514 is comprised of two subclusters: A514NW and A514SE. A514NW (A514SE) is further resolved into 
    $\mathrm{NW_{E}}$ and $\mathrm{NW_{W}}$ ($\mathrm{SE_{N}}$ and $\mathrm{SE_{S}}$).
    All four mass peaks, whose significances range from $3.4\sigma$ to $4.8\sigma$,  are centered on the brightest galaxies. The mass clump denoted as WL J0447.7-2020 is 
    a background cluster detected at the $6.1\sigma$ level.}
    \label{fig:mass map full}
\end{figure*}

\begin{figure*}
    \centering
    \includegraphics[width=0.9\columnwidth]{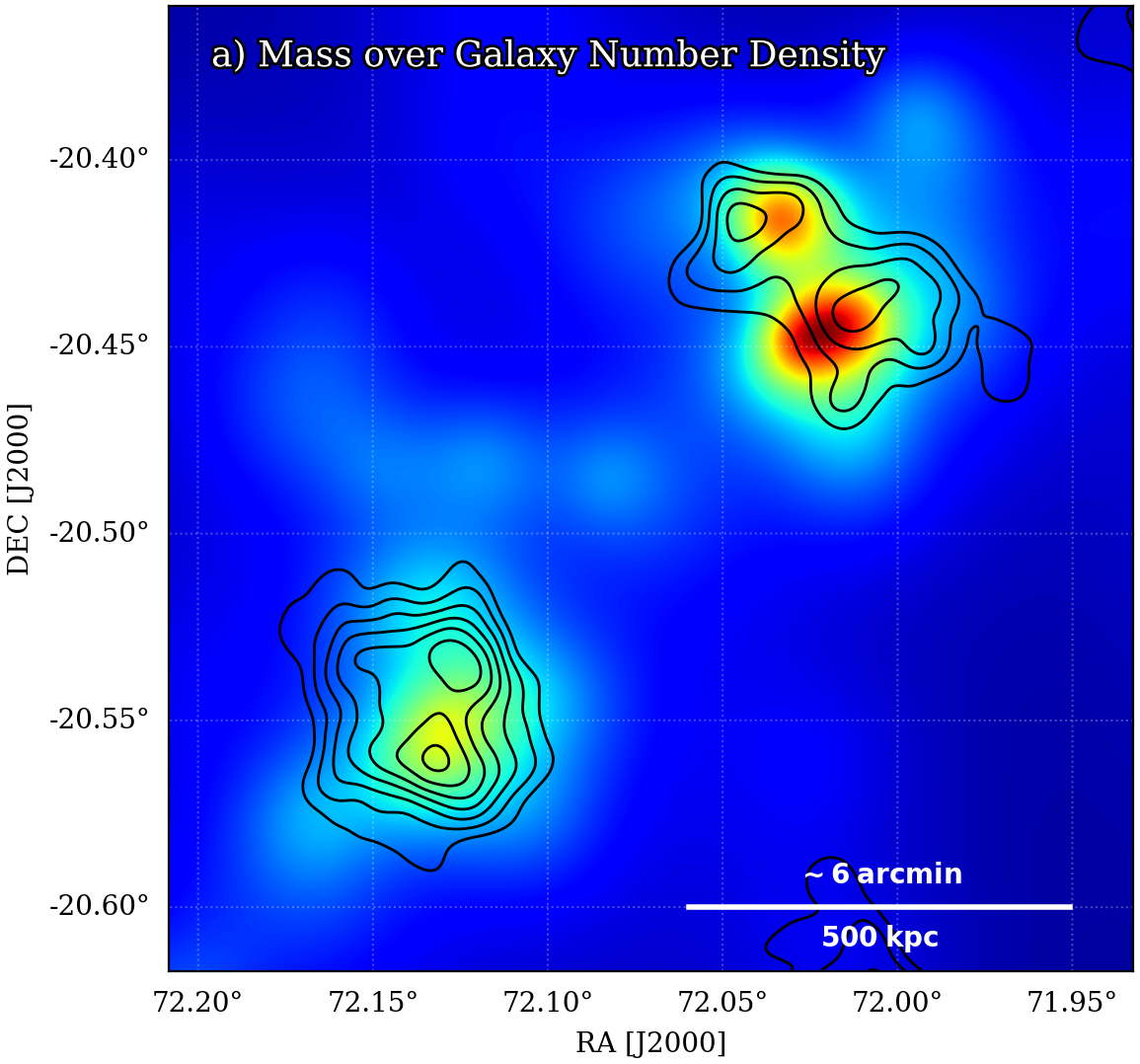}
    \includegraphics[width=0.9\columnwidth]{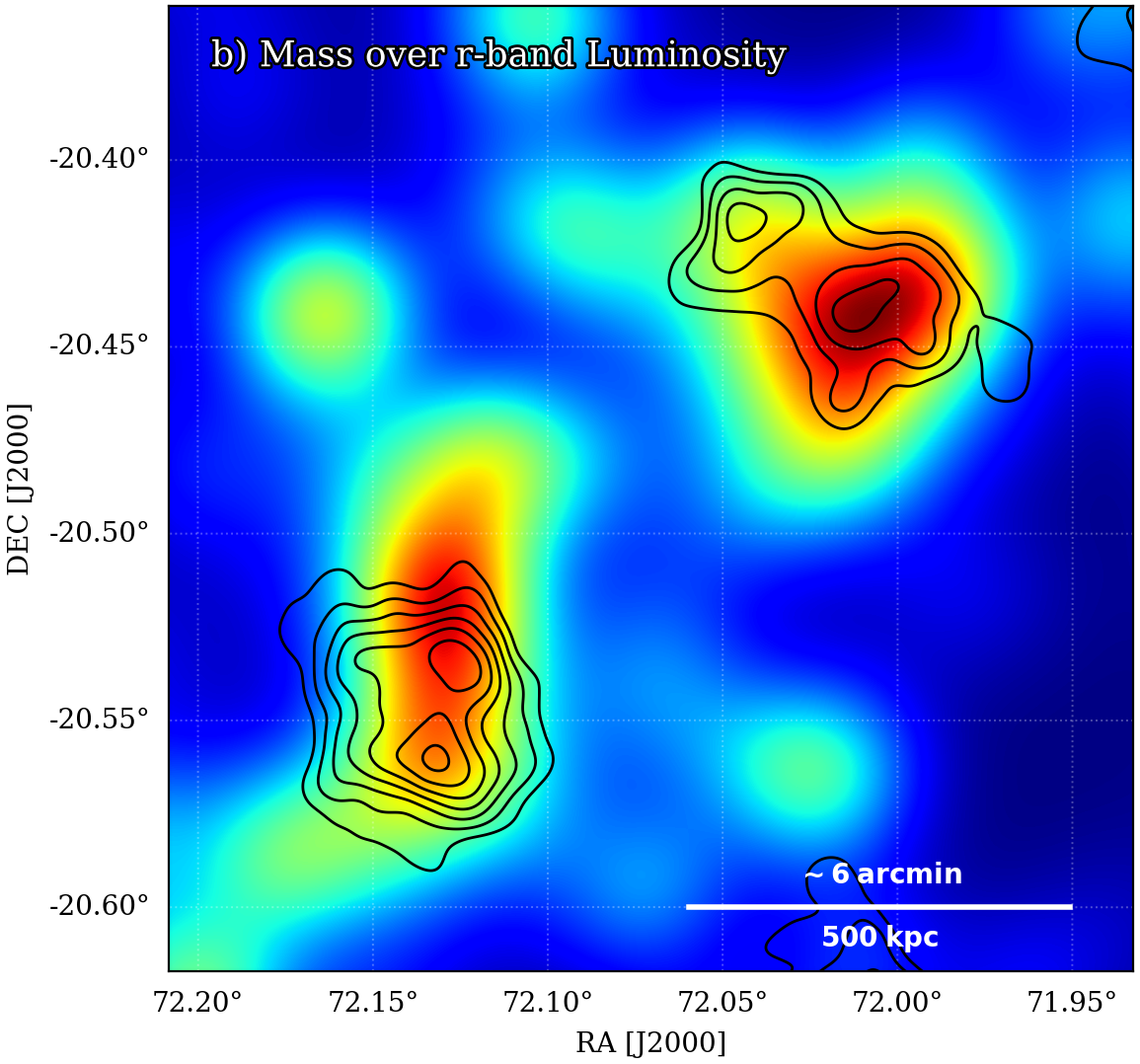}
    \includegraphics[width=0.9\columnwidth]{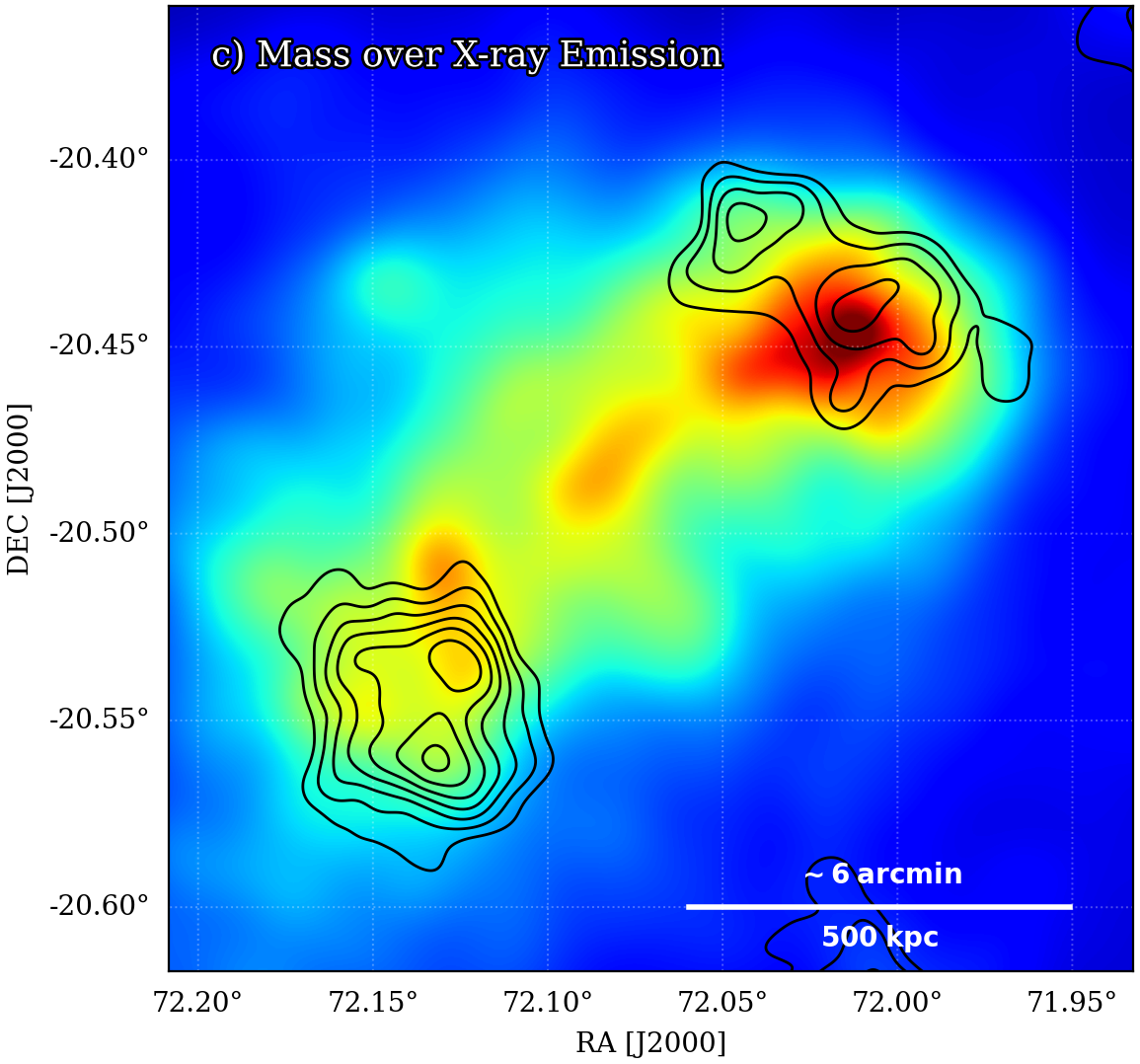}
    \includegraphics[width=0.9\columnwidth]{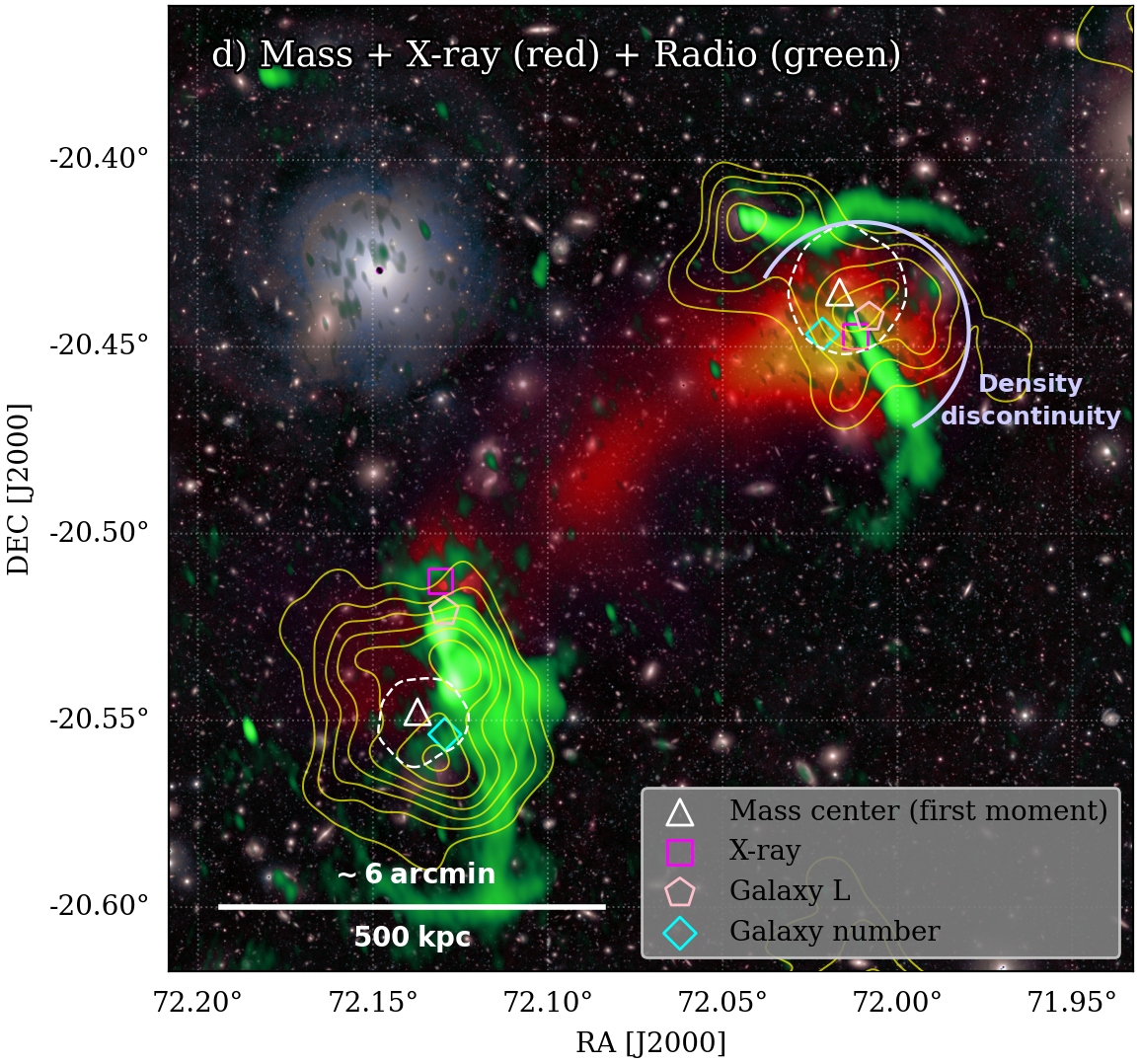}
    \caption{Comparison of the A514 mass distribution with the galaxy, X-ray, and radio distributions. a) Mass (black)  overlaid with the adaptively smoothed galaxy number density. b) Mass (black) overlaid with the galaxy luminosity in $r$-band. c) Mass (black) overlaid with the adaptively smoothed XMM-Newton X-ray emission. d) Mass (white) overlaid with the Magellan color composite + XMM-Newton X-ray (red) + uGMRT band-3 radio (green). 
    The galaxies used in panels a) and b) are the spectroscopic  and photometric (red-sequence) members. In panel d), the luminosity peaks marked with orange pentagons are estimated by flux-weighted averaging. The mass centers marked with white triangles are obtained from the first moments of the local convergence values. White dashed contours represent the $68\%$ confidence intervals of the mass centroids determined from 1000 bootstrap realizations (\textsection\ref{sub:centroid significances}).}
    \label{fig:mass map overlays}
\end{figure*}

\subsection{Mass Reconstruction}\label{sub:mass reconstruction}
One of the primary goals of our weak-lensing analysis is the robust mass reconstruction of A514 and the identification of its substructures.
Since the convergence $\kappa$ is proportional to the projected surface mass density (Eqn~\ref{eq:convergence}), hereafter we use the terms ``mass map'' and ``convergence map'' interchangeably to refer to the reconstructed convergence map.
We used the {\tt FIATMAP} code \citep{Fischer.1997, Wittman.2006}, which implements the \cite{Kaiser.1993} shear-to-convergence inversion in real space.

Figure \ref{fig:mass map full} presents the reconstructed mass distribution of the A514 field.
The overall mass distribution can be characterized by the two bimodal mass distributions: A514NW and A514SE separated by $\mathrm{\mytilde 750~ kpc}$ ($\mytilde 9'$) with A514NW (A514SE) further resolved into $\mathrm{NW_{E}}$ and $\mathrm{NW_{W}}$ ($\mathrm{SE_{N}}$ and $\mathrm{SE_{S}}$).
Our bootstrapping analysis shows that all four mass peaks have a significance greater than 3.4~$\sigma$ (see \textsection\ref{sub:centroid significances}).
We found that the mass clump denoted by WL~J0447.7-2020 is a background cluster at $z\sim0.6$.
The mass peak is in excellent agreement with its BCG candidate.
Since this cluster has not been reported in the literature, this marks its first detection based purely on WL signal.
We present additional information on this cluster in Appendix~\ref{sec:appendixA}.
We note that the positions and significances of the five mass peaks described above are also consistent with the results obtained from other mass reconstruction algorithms such as the one in \cite{Kaiser.1993}.

We label the six brightest cluster galaxies in A514 in Figure~\ref{fig:mass map full}.
Since it is difficult to obtain accurate photometry, we do not attempt to order them according to their brightness level; the suffix ``a''-``c'' is assigned arbitrarily.
All four WL peaks in A514 coincide with the brightest cluster galaxies.
Remarkably, among these four galaxies, three of them, A514SEb, A514NWa, and A514NWb, are the active galactic nuclei (AGNs) with strong bent radio jet emissions \citep{Wonki.2023b}.
The radio galaxy A514SEb, which is hosted by the mass clump $\mathrm{SE_N}$, exhibits a relatively high line-of-sight velocity difference compared to the cluster redshift and the neighboring two brightest galaxies.

Figure \ref{fig:mass map overlays} presents overlays of the mass contours with the following: a) number density, b) $r$-band luminosity, c) X-ray emission, and d) X-ray plus radio emissions.

The comparison with the galaxy number density and luminosity maps shows that the galaxy distributions correlate well with the WL mass, although scrutiny suggests that the correlations might not be as strong as those observed in the BCG-mass peak comparison mentioned earlier.
In the A514NW region, there are two number density peaks with $\mytilde1'$ offsets from their nearest mass peaks (Figure~\ref{fig:mass map overlays}a).
In contrast, the luminosity map shows only one dominant peak centered on $\mathrm{NW_{W}}$ (Figure~\ref{fig:mass map overlays}b).
In the A514SE region, the number density map shows one weak peak between $\mathrm{SE_{N}}$ and $\mathrm{SE_{S}}$.
In the luminosity map, the distribution is highly elongated in the N-S direction with its peak $\mytilde1'$ north of $\mathrm{SE_{N}}$.

The smoothed northern X-ray peak aligns well with $\mathrm{NW_{W}}$, while the southern X-ray peak is $\mytilde1'$ north of $\mathrm{SE_{N}}$ (Figure~\ref{fig:mass map overlays}c). Since the current X-ray observation is not sufficiently deep (three pointings of $\mytilde 15~$ks observation), the presence and location of the southern X-ray peak are uncertain.
However, despite the low exposure, the current X-ray data indicate significant emission between A514NE and A514SW forming a ``ridge", which serves as evidence for the post-merger status.

Figure~\ref{fig:mass map overlays}d shows that, as mentioned above, three bright bent radio jets originate from
the three brightest galaxies A514NWa, A514NWb, and A514SEb, which are hosted by the mass clumps $\mathrm{NW_{W}}$, $\mathrm{NW_{E}}$, and $\mathrm{SE_{N}}$, respectively.
The surface brightness discontinuity claimed by \cite{Weratschnig.2008} is located near the northwestern boundary of the northern X-ray emission.


\subsection{Mass Estimation}\label{sub:mass estimation}
In this section, we present mass estimation based on fitting multiple Navarro–Frenk–White \citep[NFW;][]{Navarro.1997} halos.
Although the three mass clumps (A514NW, A514SE, and WL J0447.7-2020) are widely separated ($\mytilde 9.0'$ between A514NW and A514SE, and $\mytilde 6.5'$ between A514NW and WL J0447.7-2020), we employ simultaneous fitting to minimize the influence from the neighboring structures in our three-halo fitting.
We also present the results from five-halo fitting ($\mathrm{NW_{E}}$, $\mathrm{NW_{W}}$, $\mathrm{SE_{N}}$, $\mathrm{SE_{S}}$, and WL J0447.7-2020).
Because of the degeneracy between the two parameters (concentration $c$ and scale radius $r_s$) in the NFW halo model, we used the mass-concentration ($M-c$) relation of \cite{Dutton.2014}.

For three-halo fitting, the mass center for each clump within A514 was determined by computing the first moment of the local convergence (white triangles in Figure \ref{fig:mass map overlays}d). 
We excluded background sources within $R<100''$ to minimize the influence of the substructures. 

\begin{figure}
    \centering
    \includegraphics[width=\columnwidth]{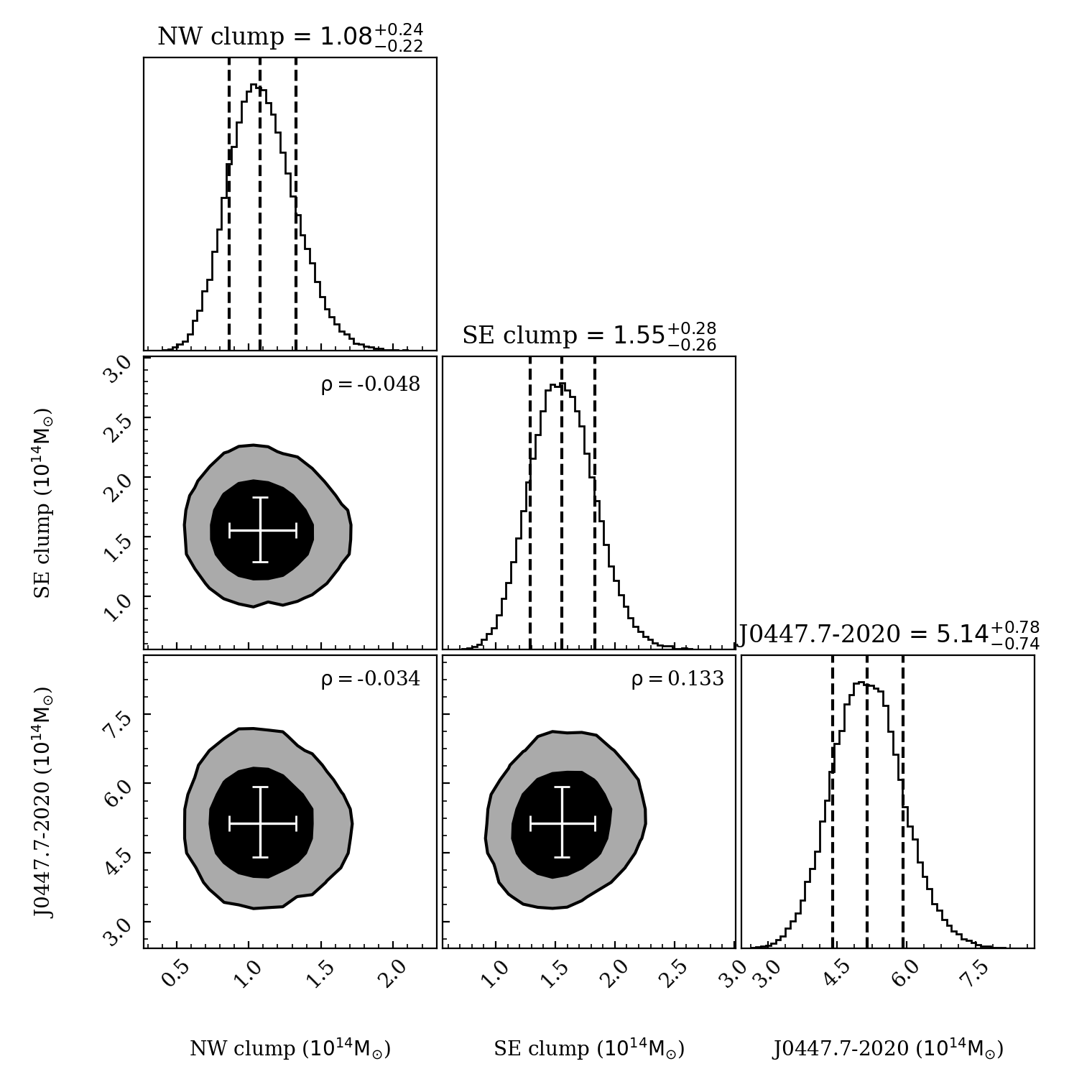}
    \caption{Posterior distributions from three-halo simultaneous mass estimation in our MCMC analysis. The mass centers were set to the first moments of the local convergence values. The black and gray filled contours represent approximately 68\% and 95\% confidence limits, respectively. The low Pearson correlation coefficients $\rho$ suggest that the degeneracies among the masses of the three halos are low.}
    \label{fig:mass estimation 3}
\end{figure}

We performed the Markov Chain Monte Carlo (MCMC) analysis
to obtain the mass posteriors with the following log-likelihood function $\mathcal{L}$ \citep{Mincheol.2019}: 
\begin{equation}
    \mathcal{L} = - \sum_{i} \sum_{j=1, 2} \frac{[g_j^m(\mathbf{M}, x_i, y_i) - g_j^o(x_i, y_i)]^2}{\sigma_{\text{SN}}^2 + (\delta e_i)^2}
\end{equation}
where $g_j^m$ is the $j$th component of the predicted reduced shear at the $i$th source galaxy position ($x_i$, $y_i$) as a function of the halo mass vector $\mathbf{M}=[M_1, M_2, ..., M_n]$, where $n$ is the number of halos, and $g_j^o$ is the $j$th component of the observed ellipticity at the same location.
We used the mathematical formulation of NFW shear \citep{Wright.2000} to calculate the $g_j^m$ at every source galaxy position.
Note that the first-order correction in Equation (\ref{eq:first-order-correction}) is applied to $g_j^m$.
We assumed a flat prior of $\mathrm{10^{12}M_{\odot}}< \mathrm{M_{200c}} < \mathrm{10^{16}M_{\odot}}$.
We display the resulting posteriors in Figure \ref{fig:mass estimation 3} (see also Table \ref{tab:mass3halo} for the summary of the best-fit results and the used mass centers).

\begin{table*}
    \centering
    \caption{Three NFW Halo Fitting Results}
    \begin{tabular}{*{7}{p{0.115\textwidth}}}
        \hline
        \hline
        \; & R.A. & Decl. & $\mathrm{M_{200c}}$ & $\mathrm{R_{200c}}$ & $\mathrm{M_{500c}}$ & $\mathrm{R_{500c}}$ \\
        \; & (J2000) & (J2000) & ($\mathrm{10^{14}M_{\odot}}$) & (Mpc) & ($\mathrm{10^{14}M_{\odot}}$) & (Mpc) \\
        \addlinespace
        \hline
        A514NW & 72.017 & -20.436 & $1.08_{-0.22}^{+0.24}$ & $0.96_{-0.07}^{+0.07}$ & $0.78_{-0.16}^{+0.17}$ & $0.64_{-0.05}^{+0.04}$ \\
        A514SE & 72.137 & -20.548 & $1.55_{-0.26}^{+0.28}$ & $1.08_{-0.06}^{+0.06}$ & $1.12_{-0.18}^{+0.20}$ & $0.72_{-0.04}^{+0.04}$ \\
        J0447.7-2020 & 71.931 & -20.340 & $5.14_{-0.74}^{+0.78}$ & $1.33_{-0.07}^{+0.06}$ & $3.52_{-0.50}^{+0.52}$ & $0.86_{-0.04}^{+0.04}$ \\
        \hline
        \label{tab:mass3halo}
    \end{tabular}
\end{table*}

\begin{figure*}
    \centering
    \includegraphics[width=0.74\textwidth]{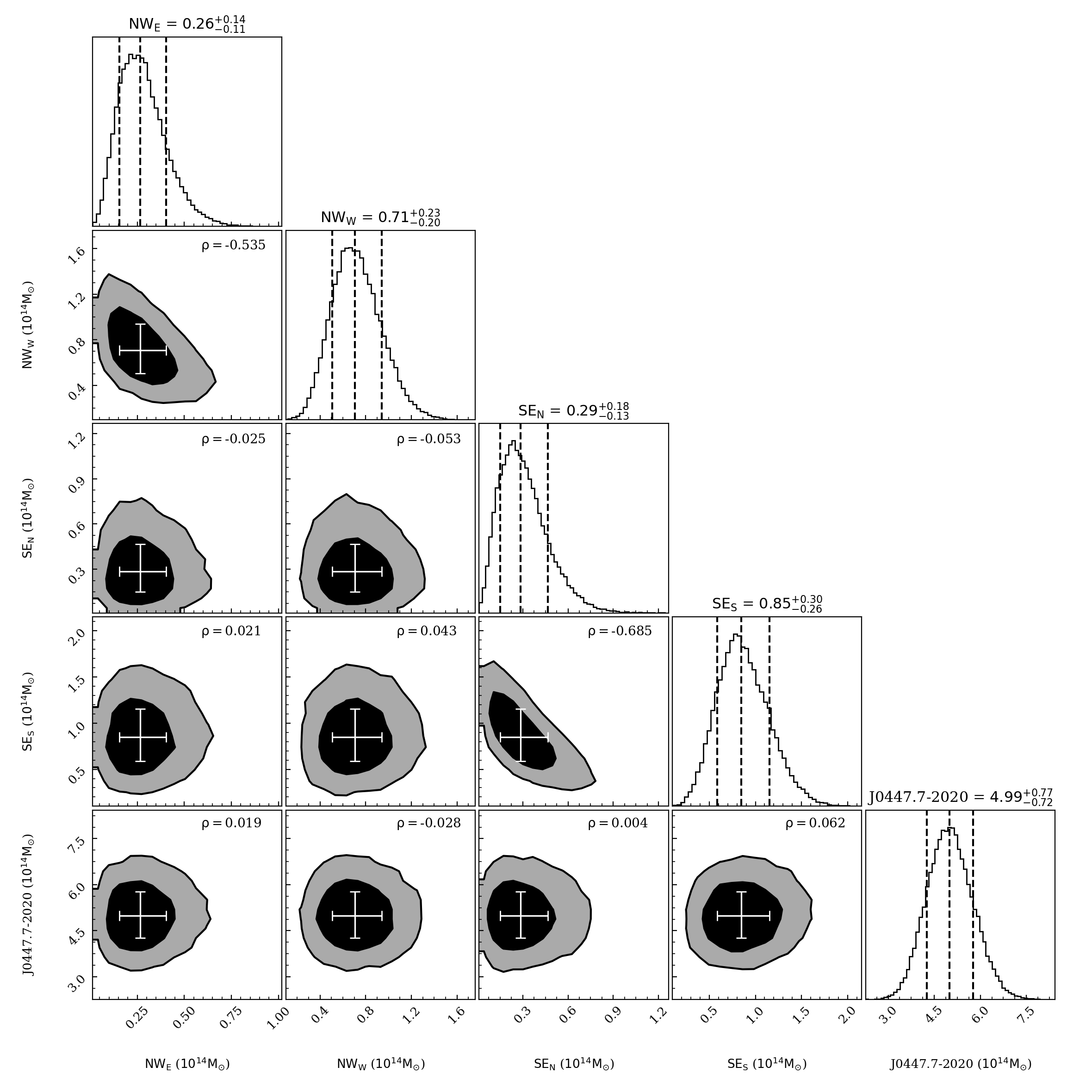}
    \caption{Five-halo mass estimation in the same format used in Figure \ref{fig:mass estimation 3}. The mass centers were set to the convergence peaks. In some cases where the distance between the two neighboring substructure is relatively small, the degeneracy is relatively high (the Pearson correlation coefficients are $\rho=-0.535$ for the $\mathrm{NW_{E} - NW_{W}}$ pair and $\rho=-0.685$ for the $\mathrm{SE_{N} - SE_{S}}$ pair).}
    \label{fig:mass estimation 5}
\end{figure*}
The masses for the NW and SE clumps are estimated to be $\mathrm{M_{200c}=1.08_{-0.22}^{+0.24} \times 10^{14} M_{\odot}}$ and $\mathrm{M_{200c}=1.55_{-0.26}^{+0.28} \times 10^{14} M_{\odot}}$, respectively.
The SE clump is marginally more massive than the NW clump.
The denser X-ray core at the NW clump (Figure~\ref{fig:mass map overlays}c) is consistent with this mass inequality because typically a less massive subcluster tends to have a more compact X-ray core in a cluster merger \cite[e.g.,][]{Clowe.2006, Mastropietro.2008, Jee.2014b, Molnar.2015, Jee.2016, Golovich.2017, Jinhyub.2021}.

\begin{table*}
    \centering
    \caption{Five NFW Halo Fitting Results}
    \begin{tabular}{*{7}{p{0.115\textwidth}}}
        \hline
        \hline
        $\;$ & R.A. & Decl. & $\mathrm{M_{200c}}$ & $\mathrm{R_{200c}}$ & $\mathrm{M_{500c}}$ & $\mathrm{R_{500c}}$ \\
        $\;$ & (J2000) & (J2000) & ($\mathrm{10^{14}M_{\odot}}$) & (Mpc) & ($\mathrm{10^{14}M_{\odot}}$) & (Mpc) \\
        \addlinespace
        \hline
        $\mathrm{NW_E}$ & 72.045 & -20.416 & $0.26_{-0.11}^{+0.14}$ & $0.60_{-0.10}^{+0.09}$ & $0.19_{-0.08}^{+0.10}$ & $0.40_{-0.07}^{+0.06}$ \\
        $\mathrm{NW_W}$ & 72.013 & -20.441 & $0.71_{-0.20}^{+0.23}$ & $0.84_{-0.09}^{+0.08}$ & $0.52_{-0.14}^{+0.16}$ & $0.55_{-0.06}^{+0.05}$ \\
        $\mathrm{SE_N}$ & 72.126 & -20.535 & $0.29_{-0.13}^{+0.18}$ & $0.62_{-0.11}^{+0.11}$ & $0.21_{-0.10}^{+0.13}$ & $0.41_{-0.07}^{+0.07}$ \\
        $\mathrm{SE_S}$ & 72.132 & -20.561 & $0.85_{-0.26}^{+0.30}$ & $0.89_{-0.10}^{+0.09}$ & $0.62_{-0.19}^{+0.21}$ & $0.59_{-0.07}^{+0.06}$ \\
        J0447.7-2020 & 71.931 & -20.340 & $4.99_{-0.72}^{+0.77}$ & $1.31_{-0.07}^{+0.06}$ & $3.42_{-0.49}^{+0.52}$ & $0.85_{-0.04}^{+0.04}$ \\
        \hline
        \label{tab:mass5halo}
    \end{tabular}
\end{table*}
The five-halo fitting results are displayed in Figure \ref{fig:mass estimation 5} and Table \ref{tab:mass5halo}.
For this analysis, the centroids of the five halos are set to the corresponding convergence peaks, which are in good agreement with the brightest cluster galaxies.
Since the separations between the two substructures in both NW and SE clumps are small ($\mytilde 2.5'$ between $\mathrm{NW_{E}}$ and $\mathrm{NW_{W}}$, and $\mytilde 1.5'$ between $\mathrm{SE_{N}}$ and $\mathrm{SE_{S}}$), the two neighboring masses are negatively correlated.

\section{Discussion}\label{sec:discussion}

\subsection{Mass Peak Significances}\label{sub:centroid significances}
Despite both the high source density and the
spatial agreement between the mass peaks and cluster galaxies, one should exercise caution against confirmation bias.
We estimated the probability that any of the five mass peaks might arise from intrinsic galaxy shape noise.
To quantify the significance, we generated 1000 convergence maps with bootstrap-resampled sources and computed the rms map.
Then, the convergence significance map was obtained by dividing the median of the 1000 bootstrap maps by the rms map.

\begin{figure}
    \includegraphics[width=\columnwidth]{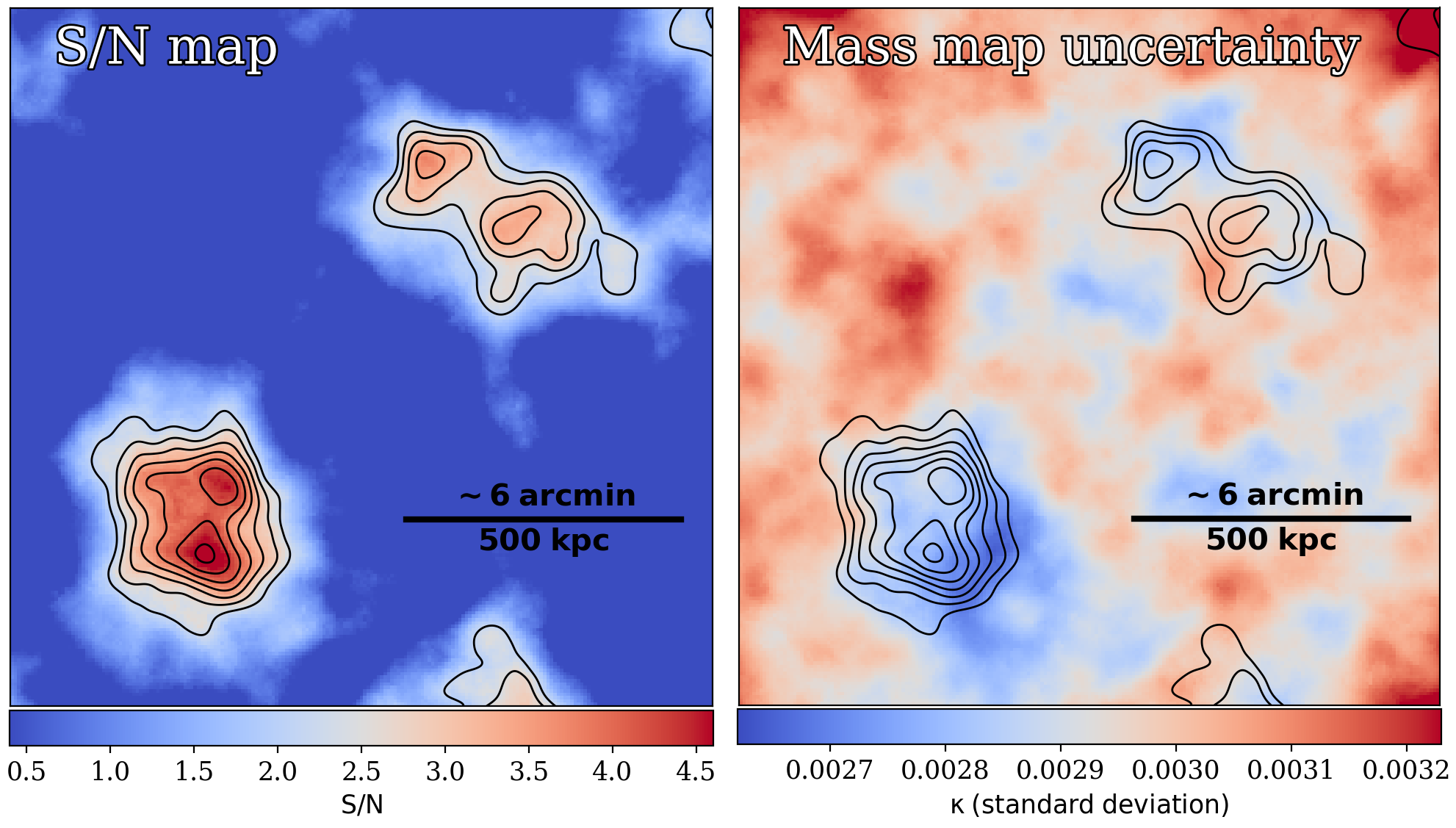}
    \caption{S/N and rms maps overlaid with mass contours. Left: Mass reconstruction S/N map obtained from 1000 bootstrapped convergence maps. $\mathrm{NW_{E}}$ and $\mathrm{NW_{W}}$ were detected at the $3.7\sigma$ and $3.4\sigma$ levels, respectively, whereas $\mathrm{SE_{N}}$ and $\mathrm{SE_{S}}$ were detected at the $4.5\sigma$ and $4.8\sigma$ levels. Right: Mass reconstruction uncertainty map. The values represent the standard deviation from the 1000 bootstrap realizations.}
    \label{fig:centroid}
\end{figure}

Figure \ref{fig:centroid} displays the significance and rms maps overlaid with the mass contours.
Both the low lensing efficiency due to the proximity of A514 and its intrinsically low mass ($\mytilde 1\mathrm{\times 10^{14} M_{\odot}}$) make the lensing signal considerably low. However, the high source density per physical area at the A514 redshift reduces the statistical noise due to the shape noise also significantly.
We find that all five substructures are detected with high significance. For the northern substructures, $\mathrm{NW_{E}}$ and $\mathrm{NW_{W}}$ are detected at the $3.7\sigma$ and $3.4\sigma$ levels, respectively.
Although $\mathrm{NW_{W}}$ is more massive than $\mathrm{NW_{E}}$, rms values are higher around $\mathrm{NW_{W}}$, so its significance is lower than $\mathrm{NW_{E}}$.
For the southern substructures, $\mathrm{SE_{N}}$ and $\mathrm{SE_{S}}$ have a significance of $4.5\sigma$ and $4.8\sigma$, respectively.
Overall rms levels are lower around the SE clump, which results in much higher significances for the southern substructures than the northern substructures.

\subsection{Uncertainties in Mass Estimation}\label{sub:uncertainties in mass estimation}
For the mass estimation (\textsection\ref{sub:mass estimation}), we assume that A514 consists of three or five spherical NFW halos.
Its mass uncertainties include only statistical noise, which comes from shape dispersion and measurement errors.

Studies have shown that systematic errors comprise a substantial fraction of the total error budget in cluster mass estimation.
For instance, departure from spherical symmetry can induce a non-negligible mass bias \citep[e.g.,][]{Clowe.2004, Feroz.2012, Limousin.2013, Herbonnet.2019}.
Moreover, WL signals can be affected by the large-scale structures (LSSs) \citep[e.g.,][]{Hoekstra.2003}.
It is difficult to remove both the LSS and triaxiality effects solely based on observational data for individual clusters.

The WL community has quantified these additional sources of mass uncertainties through numerical simulations \citep[e.g.,][]{Meneghetti.2010, Becker.2011}.
\cite{Meneghetti.2010} conducted realistic simulations of lensing observations using simulated clusters and found that the scatter of the lensing mass due to triaxiality is $\mytilde 20\%$ or higher.
\cite{Becker.2011} investigated the bias and scatter in WL mass from fitting an NFW halo to the shear profile and reported that the intrinsic scatter ranges from $\mytilde 20\%$ for massive clusters to $\mytilde 30\%$ for group-sized systems.
Since for low-mass cluster halos, the total intrinsic scatter is dominated by uncorrelated LSSs, it is plausible that for the substructures of A514, whose masses are orders of $\mathrm{10^{13} M_{\odot}}$, the scatter can be up to $\mytilde 30\%$.

Since the $M-c$ relation is the average relation of a wide variety of halos, it is reasonable to suspect that its application is not optimal for individual halos, especially for merging clusters whose profiles are significantly disrupted.
\cite{Wonki.2023a} demonstrated that the WL mass bias resulting from this model bias on post-mergers is $\lesssim \!\! 15\%$ for collisions involving A514-sized ($\mathrm{\mytilde 10^{14}M_{\odot}}$) systems.


In high-redshift clusters, the uncertainty of the source redshift distribution is a critical source of systematic errors.
However, in the current study, the low redshift of A514 makes its effect negligible.
When we vary the effective source redshift within the range $0.5<\mathrm{z_{eff}}<0.7$, the resulting change in mass is only $\mytilde4\%$.

\subsection{Possible Merging Scenarios}\label{sub:possible merging scenarios}
Although observations provide only a single snapshot of the Gyr-long merging event, a joint analysis with multi-wavelength data helps to constrain the merger stage. 
We combine the current WL result with the X-ray and radio observations to propose a merging scenario.

First of all, the $\mytilde1$~Mpc-long elongation of the X-ray emission suggests that the NW and SE subclusters have passed through each other.
In particular, the X-ray ``ridge" between NW and SE is strong evidence of the post-collision phase.
The detection of the weak shock by the density discontinuity in the NW boundary, as claimed by \cite{Weratschnig.2008}, also supports this post-collision hypothesis.
Interestingly, the density discontinuity region roughly coincides with the locations where the two northern radio jets bend.

The projected distance between the NW and SE subclusters is $\mytilde0.7$~Mpc.
Even when we assume that the merger is happening in the plane of the sky, the distance is nearly the maximum separation for a head-on collision.
When we carry out numerical simulations of head-on collisions between the two free-falling $1\times10^{14}M_{\sun}$ and $1.5\times10^{14}M_{\sun}$ halos, we find that the maximum separation after the impact reaches $\mytilde0.7$~Mpc.
The second impact happens $\mytilde0.4$~Gyr after the first apocenter.
The maximum separation after the second impact is $\mytilde0.4$~Mpc.
The presence of the aforementioned X-ray ``ridge" with the intact NW X-ray peak suggests that the collision likely occurred with a non-negligible impact parameter.
This off-axis collision is also required to explain the current $\mytilde0.7$~Mpc separation if the merger is happening not precisely in the plane of the sky (since an off-axis collision leads to a larger apocenter distance).
Non-negligible redshift differences of the brightest cluster galaxies with respect to the cluster redshift (Figure \ref{fig:mass map full}) further support the idea that the merger is not perfectly occurring in the plane of the sky.

The bimodality in both the NW and SE substructures is an interesting feature.
If those substructures are undergoing mergers within each NW or SE clump, they might have influenced the peculiar morphologies of the radio jets \citep[e.g.,][]{ZuHone.2021}.
However, with the current data, it is difficult to find any direct hint of interactions between the subclumps.
A deeper and higher-resolution X-ray observation with the {\it Chandra} observatory would be needed to address the issue.

\section{Summary}\label{sec:Summary}
A514 is a merging galaxy cluster exhibiting several intriguing multi-wavelength features.
Its X-ray emission shows a Mpc-long elongation connecting the NW and SE subclusters. 
In addition, the A514 system hosts three large-scale ($300\sim500$~kpc) bent radio jets associated with its three brightest cluster galaxies.

Our weak-lensing study with Magellan/Megacam observations revealed two mass clumps within A514 separated by $\mytilde0.7$~Mpc in the NW-SE direction.
Each mass clump is further resolved into two subclumps separated by $150\mytilde200$~kpc.
Remarkably, all four mass peaks are in good agreement with the four brightest elliptical galaxies at the cluster redshift, and three of them are the AGNs that source the large-scale bent radio jets.
In addition to the four mass peaks within A514, our WL mass reconstruction also detected a background cluster at $z\sim0.6$ located $\mytilde6\arcmin$ northwest of A514NW.

We use simultaneous NFW halo fitting with the $M-c$ relation to determine the substructure masses.
The total mass of the NW subcluster is estimated to be $\mathrm{M^{NW}_{200c} = 1.08_{-0.22}^{+0.24} \times 10^{14} M_{\odot}}$ and the mass of the SE subcluster is $\mathrm{M^{SE}_{200c} = 1.55_{-0.26}^{+0.28} \times 10^{14} M_{\odot}}$.
The masses for the four substructures are on the order of $\mathrm{10^{13}M_{\odot}}$, and their significances range from $3.4\sigma$ to $4.8\sigma$.

Both the X-ray morphology and the large ($\mytilde0.7$Mpc) separation between the NW and SE subclusters suggest that on a large scale, A514 is an off-axis post-merger system observed nearly at its apocenter. However, with the current data, it is difficult to investigate the merger phase between the two subclumps within each subcluster.

M. J. Jee acknowledges support for the current research from the National Research Foundation (NRF)
of Korea under the programs 2022R1A2C1003130 and RS-2023-00219959.

\appendix
\section{Background Cluster WL J0447.7-2020}\label{sec:appendixA}
The background cluster WL J0447.7-2020 is a massive cluster ($\mathrm{M_{200c} \mytilde 5 \times 10^{14} M_{\odot}}$) located  $\mytilde 6.5'$ northwest of A514NW. 
In principle, the presence of this massive cluster, albeit weakly, contaminates the lensing signal from A514.
We model the lensing signal from WL J0447.7-2020 to minimize its influence on the A514 mass estimation.

\begin{figure}
    \centering
    \includegraphics[width=0.85\textwidth]{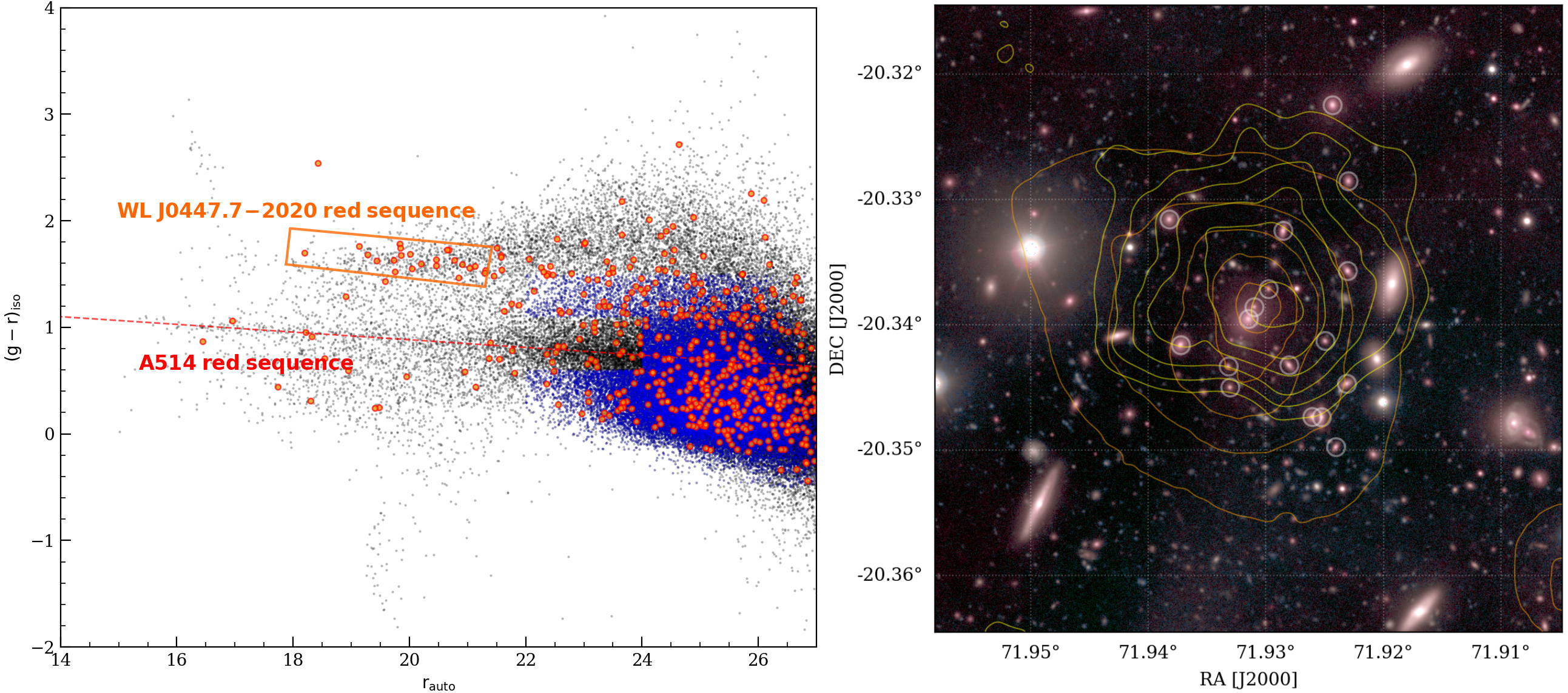}
    \caption{Left: Color-magnitude diagram of the background cluster WL J0447.7-2020. Objects located within a $r=80''$ circle around the BCG of this cluster are marked as orange circles, exhibiting a distinct red sequence redder than that of A514 (red dashed line). Blue dots represent the sources selected in \textsection\ref{sub:source selection}. Right: Mass contours (yellow) and XMM-Newton X-ray emission (orange) overlaid on the color image of WL J0447.7-2020. Cluster member candidates are marked as white circles.}
    \label{fig:background_color}
\end{figure}

In the left panel of Figure \ref{fig:background_color}, we present a color-magnitude diagram, where we mark the locations of its red-sequence candidates, which are redder than those of A514.
We selected the member galaxy candidates based on the color-magnitude relation, which are marked as white open circles in the right panel of Figure \ref{fig:background_color}.
Photometric redshifts for these candidates were measured using {\tt EAzY} \citep{Brammer.2008}, employing five PanSTARRS-1 filters ($g$, $r$, $i$, $z$, $y$) and two Megacam filters ($g$, $r$).
Given that the BCG is the dominant source among these galaxies ($\mytilde 1$ mag brighter than the second brightest galaxy) and their color-magnitude relation is tight, we utilized the {\tt CARNALL\_SFHZ\_13} models, which are suitable for quiescent galaxies.
The mean redshift for these candidates is approximately $z \mytilde 0.6$.
Therefore, we used $z=0.6$ for interpreting the shear signals from this cluster.

Applying the same method for the source redshift estimation in \textsection\ref{sub:redshift estimation}, we obtained $\left\langle \beta \right\rangle = 0.368$ and $\left\langle \beta^2 \right\rangle = 0.192$.
These values correspond to an effective source redshift of $\mathrm{z_{eff}=1.078}$, and adjust the observed shear $g'$ by a factor of $(1 + 0.418\kappa)$.

If we neglect to model the shear from this cluster, the mass of A514NW is $\mytilde5\%$ overestimated, while the mass of A514SE is $\mytilde15\%$ underestimated from the three-halo fitting results.
The photometric redshift of this cluster is currently a preliminary estimate based on seven optical and near-IR filters.
This rough estimate not only impacts the mass estimation of the cluster itself but also could affect the mass estimation of A514 because of variations in shear models depending on the cluster's redshift.
However, even if we vary the redshift of this cluster within the broad range of $0.2<z<1.0$, scatters of the masses for both A514NW and A514SE are less than $\mytilde3\%$. Thus, we conclude that the A514 mass bias due to the background cluster contamination is much smaller than its statistical error.

\bibliography{reference}

\end{document}